\documentclass[aps,prb,superscriptaddress,showpacs, preprint
]{revtex4}
\usepackage{graphicx}
\usepackage{epstopdf}
\usepackage{amssymb,amsmath}

\newcommand{\sco} {{\mbox{SeCuO${}_3$}}}
\begin{document}
\title{Site-selective quantum correlations revealed by magnetic anisotropy in the tetramer system $\sco$}
\author{I. \v Zivkovi\'c}
\affiliation{Institute of Physics, Bijeni\v cka 46, HR-10000, Zagreb, Croatia}
\author{D. M. Djoki\'c}
\affiliation{Institute of Condensed Matter Physics, EPFL, CH-1015 Lausanne, Switzerland}
\author{M. Herak}
\affiliation{Institute of Physics, Bijeni\v cka 46, HR-10000, Zagreb, Croatia}
\author{D. Paji\'c}
\affiliation{Department of Physics, Faculty of Science, Bijeni\v cka c.32, HR-10000 Zagreb, Croatia}
\affiliation{Institute of Mthematics, Physics and Mechanics, Jadranska c. 19, Sl-1000 Ljubljana, Slovenia}
\author{K. Pr\v sa}
\affiliation{Institute of Condensed Matter Physics, EPFL, CH-1015 Lausanne, Switzerland}
\author{P. Pattison}
\affiliation{ESRF, SNBL, F-38042 Grenoble 9, France}
\author{D. Dominko}
\affiliation{Institute of Physics, Bijeni\v cka 46, HR-10000, Zagreb, Croatia}
\author{Z. Mickovi\'c}
\affiliation{Institute of Condensed Matter Physics, EPFL, CH-1015 Lausanne, Switzerland}
\author{D. Cin\v ci\'c}
\affiliation{Department of Chemistry, Faculty of Science, Horvatovac 102A, HR-10000 Zagreb, Croatia}
\author{L. Forr\'{o}}
\affiliation{Institute of Condensed Matter Physics, EPFL, CH-1015 Lausanne, Switzerland}
\author{H. Berger}
\affiliation{Institute of Condensed Matter Physics, EPFL, CH-1015 Lausanne, Switzerland}
\author{H. M. R\o nnow}
\affiliation{Institute of Condensed Matter Physics, EPFL, CH-1015 Lausanne, Switzerland}

\date{\today}

\begin{abstract}
We present the investigation of a monoclinic compound SeCuO$_3$ using x-ray powder diffraction, magnetization, torque and electron-spin-resonance (ESR). Structurally based analysis suggests that SeCuO$_3$ can be considered as a 3D network of tetramers. The values of intra-tetramer exchange interactions are extracted from the temperature dependence of the susceptibility and amount to $\sim 200$ K. The inter-tetramer coupling leads to the development of long-range antiferromagnetic order at $T_N = 8$ K. An unusual temperature dependence of the effective $g$-tensors is observed, accompanied with a rotation of macroscopic magnetic axes. We explain this unique observation as due to site-selective quantum correlations.
\end{abstract}


\maketitle

%
%
%
%

\section{Introduction}
\label{Introduction}

Since the first investigations of copper(II)-acetate~\cite{Bleaney1952}, magnetic properties of localized clusters of spins have attracted considerable interest. Recent discoveries of single-molecule magnets (SMMs) like Mn$_{12}$-acetate (Ref.~\onlinecite{Sessoli1993}), V$_{15}$ (Ref.~\onlinecite{Salman2008}) and Fe$_{8}$ (Ref.~\onlinecite{Barra1996}) have revealed intriguing properties such as macroscopic quantum tunnelling of magnetization, magnetic avalanches and Berry-phase interference~\cite{Friedman2010}. Since the magnetic interaction between the individual SMMs is very weak, they can be considered as isolated entities. On the other hand, many interesting phenomena can be observed when the inter-cluster interaction is increased, allowing the system to experience a collective behavior.

Recently, a system Cu$_2$Te$_2$O$_5X_2$ ($X$ = Br, Cl) composed of Cu$^{2+}$ spins forming well-defined tetrahedra has been investigated in detail~\cite{Lemmens2001,Prester2004,Zaharko2004,Prsa2009}. Due to the antiferromagnetic (AFM) intra-tetrahedron coupling ($J \sim 40$ K), the ground state of the localized cluster is a nonmagnetic singlet. The long-range ordered state is induced in both compounds (11 K and 18 K for Br and Cl, respectively) by the presence of the inter-tetrahedra interaction ($J/T_N \approx 2 - 3$). However, it has been shown~\cite{Prsa2009} that the usual mean-field approach cannot satisfactory explain the excitation spectrum from the inelastic neutron scattering. It has been suggested that the inter-cluster quantum effects should be taken into account in a development of a new theoretical approach~\cite{Prsa2009}. In that context it is important to discriminate between the material-specific properties and a more general model.

The compound $\sco$, possessing a monoclinic space group $P2_1/n$ (Ref.~\onlinecite{Effenberger1986}), can be considered as a potential candidate for the investigation of inter-cluster quantum effects. Magnetic clusters in this compound consist of four $S = 1/2$ spins with AFM intra-cluster coupling and a weak inter-cluster interaction that leads to long-range order. In $\sco$ the spins are arranged in linear segments - tetramers. We present the evidence that the intra-tetramer interactions are at the order of 200 K while the ordering takes place at $T_N = 8$ K. This implies that the temperature range where the effects of inter-cluster quantum fluctuations could be investigated is considerably larger ($J/T_N \approx 20$) compared to Cu$_2$Te$_2$O$_5X_2$.

We have also discovered a unique influence of quantum correlations on the magnetic anisotropy of the system. The torque magnetometry revealed that the macroscopic magnetic axes are rotating with temperature and the ESR measurements showed a strong temperature dependence of the effective $g$-factor along specific crystallographic directions. We show that both observations can be explained as a consequence of the singlet formation of the central copper pair within the tetramer.

%
%
%
%

\section{Experimental details}
\label{Details}

Single crystals of $\sco$ have been grown by a standard chemical vapor phase method. Mixtures of analytical grade purity CuO and SeO$_2$ powder in molar ratio 4 : 3 were sealed in the quartz tubes with electronic grade HCl as the transport gas for the crystal growth. The ampoules were then placed horizontally into a tubular two-zone furnace and heated very slowly by 50 $^0$C/h to 500 $^0$C. The optimum temperatures at the source and deposition zones for the growth of single crystals have been 550 $^0$C and 450 $^0$C, respectively. After four weeks, many green $\sco$ crystals with a maximum size of 5x10x2 mm$^3$ were obtained, which were identified on the base of X-ray powder diffraction data.

The temperature dependent powder diffraction data were collected on beamline BM01A at the Swiss-Norwegian Beamline (SNBL) in the ESRF, Grenoble. A monochromatic wavelength of 0.6971 \AA~was selected, and the beam focused onto the sample using a combination of curved mirror and monochromator crystal. The sample of powdered single crystal of $\sco$ was loaded into a 0.3 mm diameter capillary. The experiment has been carried out on a MAR-Research mar345 image plate and the sample was rotated 30$^0$ during an exposure of 30s per frame. The temperature was controlled by a cryostream N2 gas flow blower from Oxford Cryosystems.

The magnetic susceptibility of a powder was measured by a Faraday method in a home-built setup in applied dc magnetic field of 5 kOe. The measurements on single crystals were performed with a Quantum Design MPMS magnetometer.

Magnetic torque was measured with a home-built torque apparatus which uses a torsion of a thin quartz fiber for the torque measurement. The sample holder is made of an ultra pure quartz and has an absolute resolution of 10$^{-4}$ dyn cm.

Temperature dependent ESR spectra (5 –- 300 K) were recorded using an ESR spectrometer, Model E540 EleXys (Bruker BioSpin GmbH), operating in the microwave X-band ($9.4$ GHz), and equipped with a cylindrical $TE_{011}$ high-Q cavity. A continuous-flow helium cryostat, Oxford Instruments Model ESR900, was connected. The spectrometer operates in reflection mode and modulation of the magnetic field $H_{0}$ was used for the signal enhancement. The $100$ kHz modulation amplitude was kept at $10$ G to avoid modulation broadening. The angular dependence was obtained using a one-axis goniometer.

%
%
%
%

\section{Experimental results}
\label{Results}

\subsection{Structure}
\label{sec-structure}

\begin{figure}
\includegraphics[width=0.45\textwidth]{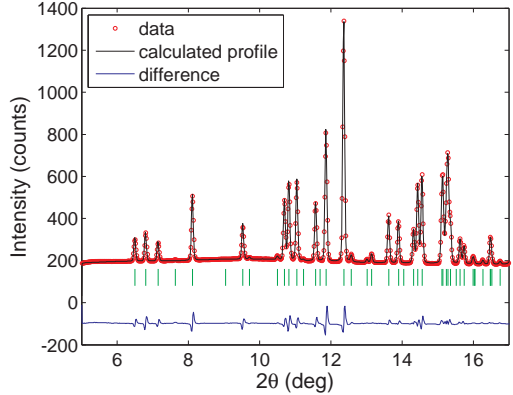}
\caption{(Color online) The X-ray powder pattern of $\sco$. The vertical markers correspond to the Bragg positions.}
\label{fig-powder}
\end{figure}

In Fig.~\ref{fig-powder} we present the powder diffraction pattern of $\sco$. There exist several compounds with the same $\sco$ chemical formula but with different crystal structures. The observed powder pattern could be indexed to a monoclinic unit cell, space group $P2_1/n$, with unit cell parameters $a = 7.712$ \AA, $b = 8.238$ \AA, $c = 8.498$ \AA~and $\beta = 99.124\ \deg$. Details of this structure have been published in Ref.~\onlinecite{Effenberger1986} among other polymorphs and the structure has been designated with Cu(SeO$_3$)-III.

\begin{figure}[h!]
\includegraphics[width=0.45\textwidth]{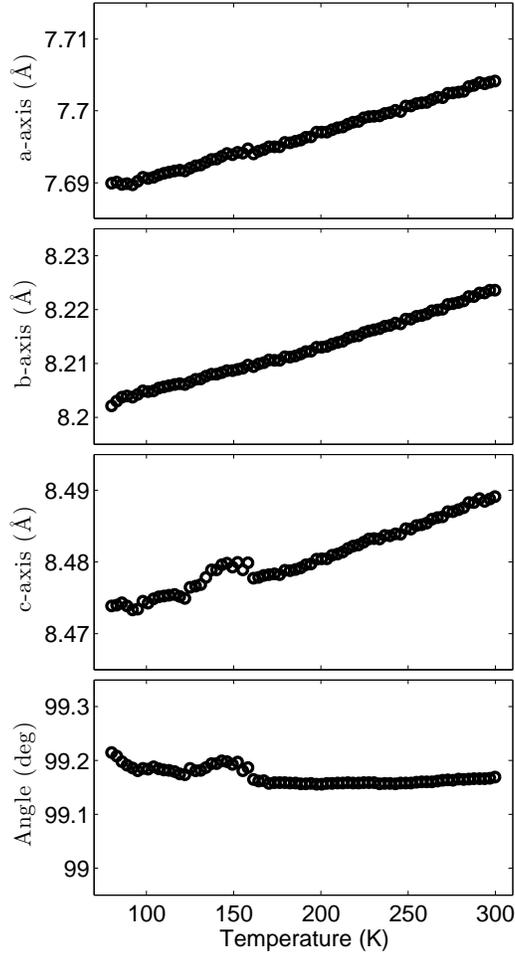}
\caption{The temperature dependence of unit cell parameters for the monoclinic $\sco$.}
\label{fig-unitcell}
\end{figure}

The temperature dependence of the unit cell parameters does not reveal any kind of structural phase transition down to 80 K (see Fig.~\ref{fig-unitcell}). We attribute the small deviation from linearity seen for the $c$-axis and the $\beta$-angle around 150 K as an experimental artefact.

\begin{figure}
\includegraphics[width=0.45\textwidth]{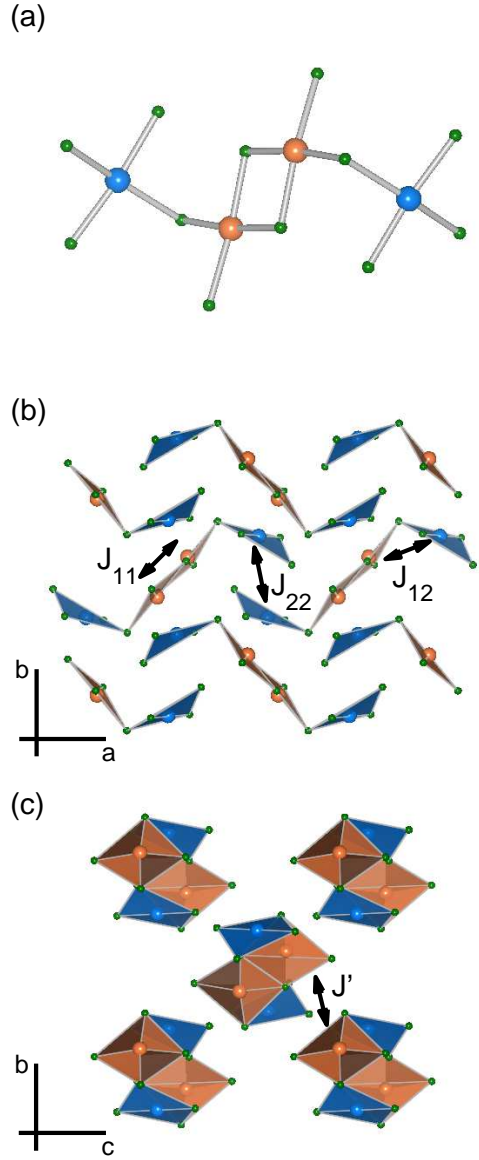}
\caption{(Color online) The sketch of the crystal structure of $\sco$ taking into account active $d_{x^2-y^2}$ orbitals~\cite{Momma2008}. (a) a single tetramer; view along the (b) $c$-axis and (c) $a$-axis. (Cu1 -- orange, Cu2 -- blue, O -- green)}
\label{fig-structure}
\end{figure}

The monoclinic SeCuO$_3$ has 2 distinct copper sites, Cu1 and Cu2. Each site is composed of a central copper ion surrounded by 6 oxygen ions, forming a distorted octahedron, Fig.~\ref{fig-structure}. Selenium ions are placed in the center of a tetrahedron formed by 3 oxygens and a lone electron pair. Lone electron pairs act as chemical scissors, effectively reducing the number of magnetic exchange paths and forming low-dimensional magnetic subsystems.

In addressing the possible influence of the crystal structure on magnetic properties, it is important to consider how the local environment around the magnetic ion affects the energy levels of $d$-orbitals. With a $d^9$ configuration on the copper ion, only the highest lying orbitals are magnetically active. In the (ideal) octahedral crystal field $d$-orbitals are split into lower lying $t_{2g}$ and higher lying $e_g$ subsets. $e_{g}$ consists of a planar $d_{x^2-y^2}$ and an elongated $d_{3z^2-r^2}$ orbitals. From the list of all the distances between the copper and oxygen ions presented in Table~\ref{Cu-O} one can immediately notice that each copper ion has 4 oxygen ions at distances around 2 \AA~and two oxygen ions with distances much larger than 2 \AA . Particularly, these two oxygen ions (O1 and O4 for the Cu1 ion and O4 and O5 for the Cu2 ion) are located on the opposite sides of the octahedron, making it elongated along the local $z$-direction. This removes the degeneracy of the $e_g$ subset in a way that leaves only $d_{x^2-y^2}$ as a highest lying orbital.
\begin{table}
\caption{Copper - oxygen distances (in \AA).}
\label{Cu-O}
\begin{tabular}{|c|c||c|c|}
\hline
Atoms &  $d$(Cu1--O) & Atoms &  $d$(Cu2--O) \\
\hline
Cu1--O1 & 2.38 & Cu2--O1 & 1.98 \\
Cu1--O3 & 1.95 & Cu2--O2 & 1.96 \\
Cu1--O3 & 1.96 & Cu2--O4 & 1.95 \\
Cu1--O4 & 2.42 & Cu2--O4 & 2.71 \\
Cu1--O5 & 2.02 & Cu2--O5 & 2.36 \\
Cu1--O6 & 1.95 & Cu2--O6 & 1.98 \\
\hline
\end{tabular}
\end{table}

The superexchange interaction Cu--O--Cu between the magnetic moments should be strongest along the path $d_{x^2-y^2}$ -- $p_{x(y)}$ -- $d_{x^2-y^2}$. If only these paths are taken into account, then the structure can be represented as a weakly-coupled 3D network of tetramers, presented in Fig.~\ref{fig-structure} with views along the $c$- and $a$-axis. Tetramers are effectively forming two sets of chains running along the $a$-axis, with tetramers in the neighboring chains related to each other by the 180$^0$ rotation around the $b$-axis.

A single tetramer consists of a central Cu1--Cu1 dimer and two Cu2 ions on each end of the dimer. There are two Cu1--O--Cu1 paths with angles of 101.9$^0$ which contribute to the $J_{11}$ interaction. On the other hand, there is a single Cu1--O--Cu2 path with an angle of 108.5$^0$ for the $J_{12}$ interaction. Since these angles are significantly above 90$^0$, one would expect medium strength AFM interactions between the neighboring moments.

Two types of inter-tetramer interactions can be expected. One is along the $a$-axis between two Cu2 ions on the neighboring tetramers, $J_{22}$, and incorporates the $d_{3z^2-r^2}$ orbital. Since the octahedron is not ideal, there maybe some mixing between $d_{3z^2-r^2}$ and $d_{x^2-y^2}$ orbitals, giving rise to a finite $J_{22}$. The second inter-tetramer interaction is the inter-chain interaction $J'$ and it is mediated via Cu--O--Se--O--Cu paths.

\subsection{Magnetization}
\label{sec-magnetization}

\begin{figure}[b]
\includegraphics[width=0.45\textwidth]{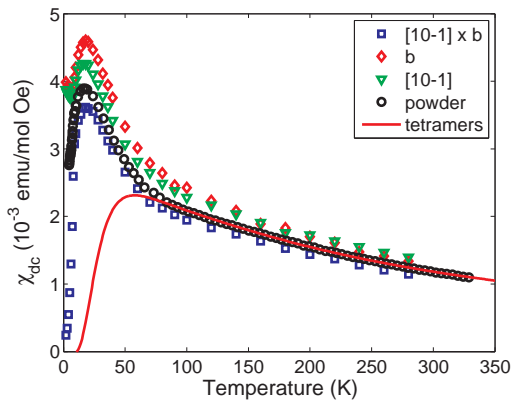}
\caption{(Color online) dc susceptibility vs. temperature for the powdered sample and three orthogonal directions of the single crystal. The solid line is obtained using Eq.~(\ref{eq-chi})}
\label{fig-susc}
\end{figure}

\begin{figure}
\includegraphics[width=0.45\textwidth]{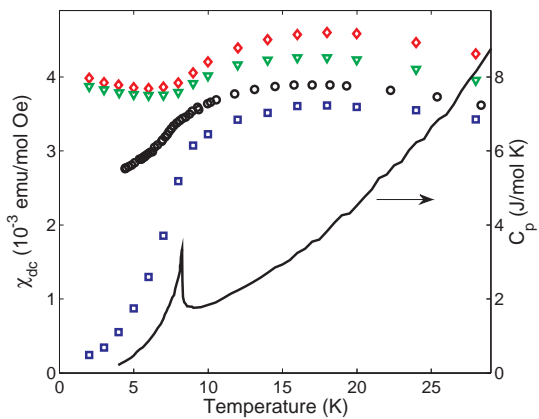}
\caption{(Color online) dc susceptibility around the transition. The markers are the same as in Fig.~\ref{fig-susc}. The line shows the temperature dependence of the specific heat.}
\label{fig-transition}
\end{figure}

The temperature dependence of the dc susceptibility $\chi_{dc} = M/H$ is presented in Fig.~\ref{fig-susc} for three different crystallographic directions and a powdered sample. Over the whole temperature range the susceptibility from the powdered sample follows the behavior expected from a direction-averaged measurement. At $T_N = 8$ K the transition to the long-range ordered state occurs (see Fig.~\ref{fig-transition}).

The dominant feature of the magnetic response of the $\sco$ compound is a broad maximum located around 18 K. The presence of such a maximum is usually related to the build up of AFM correlations in the absence of long-range order due to the strong fluctuations present in low-dimensional systems. Following our structural analysis, the hamiltonian that describes isolated tetramers is given by
\begin{equation}
{\cal H} = J_{12}(\boldsymbol{S_1} \cdot \boldsymbol{S_2} + \boldsymbol{S_3} \cdot \boldsymbol{S_4}) + J_{11}(\boldsymbol{S_2} \cdot \boldsymbol{S_3}).
\label{eq-Hamil}
\end{equation}
where $\boldsymbol{S_2}$ and $\boldsymbol{S_3}$ form the central pair coupled with the $J_{11}$ exchange interaction and end spins are $\boldsymbol{S_1}$ and $\boldsymbol{S_4}$ coupled to the central pair with $J_{12}$.

Similar to the case of isolated tetrahedra in Cu$_2$Te$_2$O$_5X_2$, for $J_{11},J_{12} > 0$ the ground state is a nonmagnetic singlet. The temperature dependence of the susceptibility is given by~\cite{Emori1975}
\begin{equation}
\chi = \frac{N_A g_{av}^2 \mu_B ^2 \beta}{2}\frac{{5exp(\beta J(2+\gamma )) + exp(\beta J \gamma ) + exp(\beta J K_+) + exp(\beta J K_-)}}{Z} + \chi_0
\label{eq-chi}
\end{equation}
where
\begin{equation}
\nonumber
Z = {5exp(\beta J(2+\gamma )) + 3exp(\beta J \gamma ) + 3exp(\beta J K_+) + 3exp(\beta J K_-) + 2cosh(\beta J K^*)}
\end{equation}
and
\begin{gather}
\nonumber
\beta = 1/k_B T \text{,} \quad J = J_{12}/2 \text{,} \quad \gamma = J_{11}/J_{12} \\
\nonumber
K_{\pm} = 1 \pm \sqrt{1+\gamma ^2} \text{,} \quad K^* = \sqrt{4-2\gamma + \gamma ^2}.
\end{gather}
Here $N_A$, $g_{av}$ and $\mu_B$ are Avogadro's number, average $g$-factor and Bohr magneton, respectively. $\chi_0$ is a temperature independent susceptibility. With $\gamma = 0$ one can recover the Bleaney-Bowers equation for a dimer system.~\cite{Bleaney1952}

As it turns out, it is not possible to model the experimental data with Eq.~(\ref{eq-chi}) in the whole temperature range using a single set of parameters. Moreover, our attempts to model it as a simpler low-dimensional system, like a 1D magnetic chain or a magnetic dimer, also could not explain the magnetic behavior of $\sco$. We believe that this is due to the additional interaction in the system, not taken into account in the tetramer hamiltonian, Eq.~(\ref{eq-Hamil}). This is reflected in the measured susceptibility as a kink around 70 K below which the susceptibility starts to grow faster. For 90 K $< T <$ 330 K one can successfully implement Eq.~(\ref{eq-chi}) with parameters $J_{11} = 225$ K, $J_{12} = 160$ K, $g_{av} = 2.25$ and $\chi_0 = -4 \cdot 10^{-5}$ emu/mol Oe. Here it is important to emphasize that neither a 1D magnetic chain model nor a magnetic dimer model can reproduce the susceptibility in this temperature range. The nature of the interaction that sets in around 70 K is unclear at the moment.

In Fig.~\ref{fig-transition} we present the results around the transition to the ordered state. All the curves exhibit a maximum around 18 K and at $T_N$ = 8 K one of them (along the $\tau = [10\bar{1}] \times b$ direction) drops sharply towards zero. The susceptibility along two other crystallographic directions remains at the same level with a slight increase towards $T = 0$ K. This behavior is usually observed in systems with an uniaxial AFM arrangement of the magnetic moments where the susceptibility along the easy-axis is greatly reduced compared to other two directions. Additionally, the specific heat measurement reveals a $\lambda$-peak, supporting the conclusion that a 3D magnetic ordering takes place.

\begin{figure}
\includegraphics[width=0.45\textwidth]{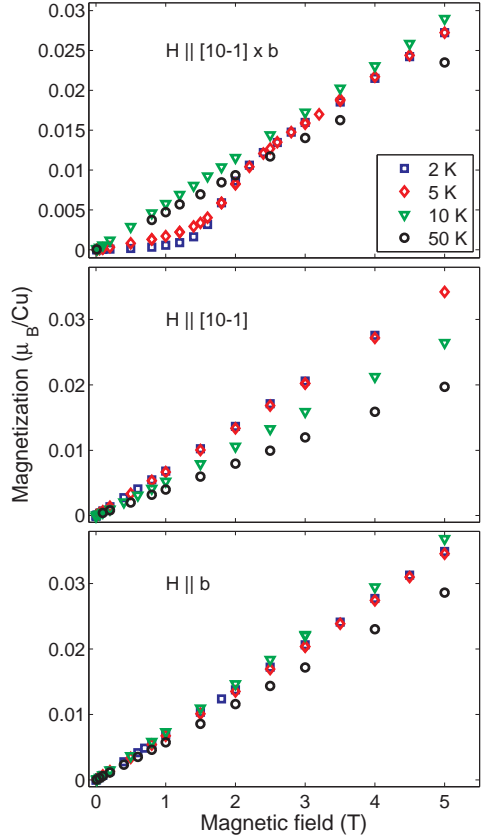}
\caption{Magnetic field scans along three crystallographic directions.}
\label{fig-MH}
\end{figure}

The AFM nature of magnetic order is further corroborated by the magnetization vs.~field scans along the three crystallographic directions, Fig.~\ref{fig-MH}. The scans with $H \perp \tau$ show a linear dependence of the magnetization at all temperatures. On the other hand, the scans along the $\tau$-direction are linear in the whole field range down to 10 K but below $T_N$ there is a deviation from the linearity when the field is smaller than 2 T. This is a characteristic behavior for the spin-flop transition where the moments switch their orientation from that parallel to the field to the one that is perpendicular to the field but still keeping the AFM alignment. The value of the spin-flop field is $\sim 1.7$ T, similar to another case with the AFM ordering of Cu$^{2+}$ spins in a 1D alternating chain CuSb$_2$O$_6$~(Ref.~\onlinecite{Heinrich2003}).

\subsection{Torque magnetometry}
\label{sec-torque}

\begin{figure}
\includegraphics[width=0.45\textwidth]{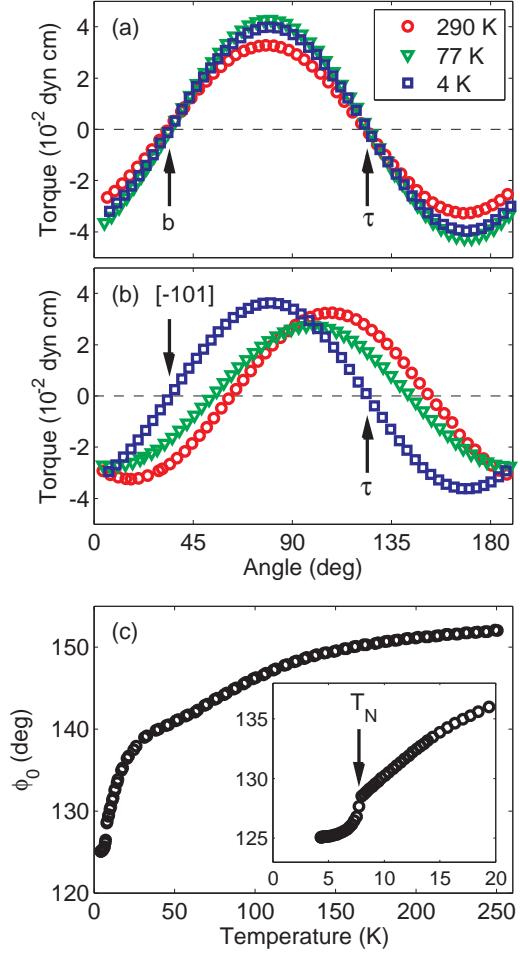}
\caption{Torque measurements in (a) $b\tau$-plane and (b) $ac$-plane. (c) Temperature dependence of the phase shift of magnetic axes in the $ac$-plane.}
\label{fig-torque}
\end{figure}

A subtle feature can be noticed around 200 K in Fig.~\ref{fig-susc}. Namely, single crystal susceptibilities along $b$- and $[10\bar{1}]$-directions cross each other which indicates that the magnetic anisotropy of the system is changing with temperature. In order to investigate this feature more closely, we have performed torque measurements within two perpendicular planes: (010) plane ($ac$-plane) and the one containing the $b$-axis and the $\tau$ direction ($b\tau$-plane). The angular dependence of the measured torque component $\Gamma_z$ in the case of a linear response is given by the expression:
\begin{equation}
\Gamma_z = \frac{m}{2M}H^2\Delta \chi_{xy} \sin (2\phi - 2\phi_0)
\label{eq-torque}
\end{equation}
where $m$ is the mass of the sample, $M$ the molar mass, $H$ the magnetic field and $\Delta \chi_{xy}=\chi_x-\chi_y$ is the susceptibility anisotropy in the $xy$ plane (plane of rotation of the magnetic field). $\phi_0$ is the phase shift with respect to the laboratory frame. From Eq.~(\ref{eq-torque}) we see that the torque is zero when the field is applied along $\phi_0$ and $\phi_0 + 90^0$. Also, the torque is a sine curve with a positive amplitude if $\chi_x>\chi_y$. The measured angular dependence of torque in $\sco$ in $b\tau$- and $ac$-planes at different temperatures is shown in Fig.~\ref{fig-torque}. In both the paramagnetic and the AFM state the measured torque obeys the above expression which means that the response to the magnetic field is linear. The angle in Fig.~\ref{fig-torque} is the goniometer angle.

One can immediately notice that for the $b\tau$-plane the zeros of the curves do not change as the temperature is lowered from the room temperature down to 4 K. In contrast, there is a substantial shift of magnetic axes within the $ac$-plane. The temperature dependence of the phase shift is presented in Fig.~\ref{fig-torque}c with the inset showing the region around the transition. At 8 K there is a clear indication of the transition, and for temperatures below 5 K the easy axis approaches the $\tau$ direction, in accordance with the magnetization results.

\subsection{Electron Spin Resonance}
\label{sec-esr}

\begin{figure}
\includegraphics[width=0.45\textwidth]{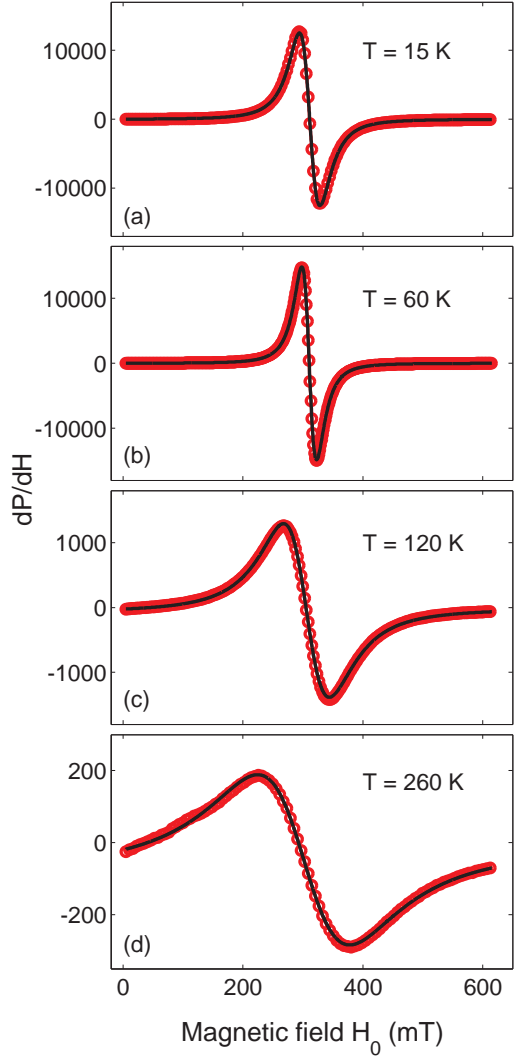}
\caption{X band (9.4 GHz) ESR absorption spectra for $H_0 \parallel [10\bar{1}]$ (red) fitted with first derivative Lorentzian lines (black).}
\label{fig-esr2}
\end{figure}

To further investigate the anisotropy, electron spin resonance measurements have been performed. The absorption profile of spectra obtained for various crystal orientations with respect to both sweeping $H_{0}$ and microwave $H_{1}$ magnetic field show a single, exchange-narrowed Lorentzian absorption line~\cite{Anderson1953,Anderson1954}. In Fig.~\ref{fig-esr2} we present several spectra taken at different temperatures for the $H_0 \parallel [10\bar{1}]$ orientation. At room temperature we have observed a slight departure of the ESR lineshape from the Lorentzian absorption profile which is explained by the overlap of tails of the ESR line counterpart appearing in the negative magnetic field sector. The original line becomes so broad as to overlap via $H_{0} = 0$ T with its negative magnetic field equivalent.

Below $T_N$ no AFM resonance absorption spectrum has been noticed to emerge at $H_{0}$ fields swept out from $50$ up to $10000$ Gauss for different crystal orientations.

\begin{figure}
\centering\includegraphics[width=0.45\textwidth]{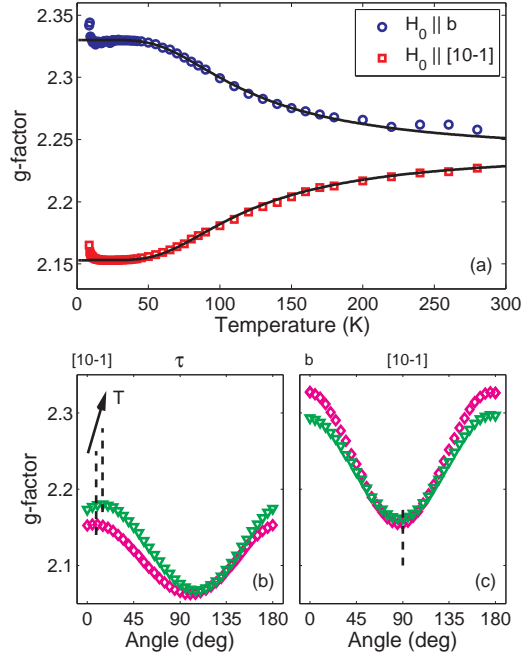}
\caption{(a) Temperature dependence of the $g$-factor with respect to two crystal orientations. The lines represent the fit using Eq.~(\ref{eq-average}). Angular dependence of the $g$-factor at 15 K ($\diamond$) and 100 K ($\triangledown$) (b) in the $ac$-plane and (c) in the $b[10\bar{1}]$-plane. Vertical dashed lines mark the position of extrema.}
\label{fig-gfactor}
\end{figure}

We have extracted the values of the $g$-factor for each absorption spectrum and in Fig.~\ref{fig-gfactor} we present its temperature and angle dependence. Above $\sim 200$ K the $g$-factors saturate to values $2.23$ and $2.26$ for $H_0 \parallel [10\bar{1}]$ and $H_0 \parallel b$, respectively. Below 200 K, however, one discerns two contrary monotonic trends down to 40 K. Between 40 K and 15 K the $g$-factors retain almost constant values and in the vicinity of $T_N$ both $g$-factors are increasing. Such a strong temperature dependence of the $g$-factor is usually associated with underlying structural changes, as evidenced in CuSb$_2$O$_6$ which exhibits a monoclinic-to-tetragonal phase transition below 400 K~\cite{Heinrich2003}.

The angular dependence at 15 K and 100 K is presented in Fig.~\ref{fig-gfactor}(b) and (c) for the $ac$- and $b[10\bar{1}]$-plane, respectively. We have marked the position of the extrema for each curve and one can immediately notice that there is a clear phase shift in the $ac$-plane, confirming the results from the torque magnetometry. On the other hand, when the anisotropy is measured in the plane that contains the $b$-axis, there is no phase shift within the experimental error.

\begin{figure}
\includegraphics[width=0.45\textwidth]{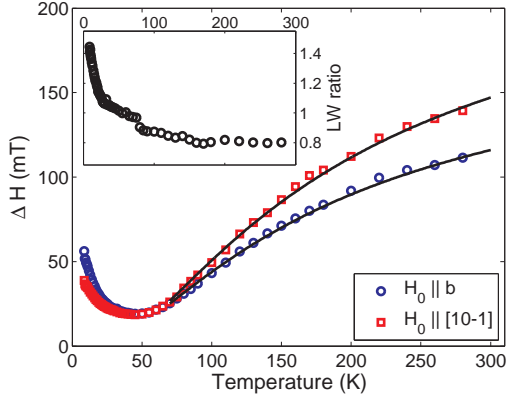}
\caption{Temperature dependence of $9.4$ GHz ESR line-widths obtained for two crystal orientations ($A(\protect\phi =0)$ and $B(\protect\phi =0)$). The data are fitted (solid lines) using Eq.~(\ref{eq-delta}). The inset shows the ratio of two line-widths.}
\label{fig-LW}
\end{figure}

In Fig.~\ref{fig-LW} we present the temperature dependence of the line-width $\Delta H$ for two crystal orientations. The line-width is very broad and reaches nearly $150$ mT at high temperatures. As the temperature is lowered, the line narrows with a broad minimum around 50 K below which $\Delta H$ tends to increase in the vicinity of the phase transition. It is interesting to note that at high temperatures $\Delta H^{[10\bar{1}]} > \Delta H^b$ while below the minimum $\Delta H^{[10\bar{1}]} < \Delta H^b$.

We can disregard the spin-diffusion as a relaxation mechanism since no deviation from the Lorentzian profile has been noticed.~\cite{Richards1976,Huber2003,Cheung1978} For classical 3D antiferromagnets ($T_{N}\sim J$) the Kubo-Tomita (KT) model\cite{Kubo1954} describes well the broadening of the absorption lines in the paramagnetic regime. However, KT model is not applicable for the low-dimensional systems where long-range order is suppressed due to the AFM correlations in the range $T_N < T < J$. More recently Oshikawa and Affleck (OA) have developed a model to cover the deficiency of the KT theory at low $T$ in $S=1/2$ AFM Heisenberg chains \cite{Oshikawa2002}. Since $\sco$ has two distinct copper site, neither the OA model can be applied.

In a recent ESR investigation of the spin-Peierls compound CuGeO$_3$ very similar behavior of the ESR line-width has been reported~\cite{Eremina2003}: a significant decrease of the line-width as the temperature is decreased with a broad minimum and a subsequent increase just above the transition. In an attempt to describe the high temperature behavior of the line-width, Eremina and coworkers used a semi-phenomenological approach~\cite{Eremina2003}
\begin{equation}
\Delta H(T)=\Delta H(\infty )e^{-\dfrac{C_{1}}{C_{2}+T}}\text{.}
\label{eq-delta}
\end{equation}
Here, $C_{1}$ corresponds to the exchange constant $J$, while $C_{2}$ is related to a transition temperature $T_N$ and is therefore associated with the inter-tetramer interaction. The parameters extracted from the fit (see Table~\ref{tab1}) have reasonable values: $C_1 \approx 170$ K, close to $J_{12}$ deduced from the tetramer fit, and $C_2 \approx 10$ K, close to the value of the 3D ordering.
\begin{table}
\caption{The fitting parameters for the line-width behavior.}
\label{tab1}\centering
\begin{tabular}{cccc}
\hline\hline
Orientation & $C_{1}$[K] & $C_{2}$[K] & $\Delta H(\infty )$[mT] \\%
[0.5ex] \hline
$A(\phi =0)$ & $175\pm 10$ & $7\pm 3$ & $260\pm 10$ \\ 
$B(\phi =0)$ & $170\pm 10$ & $12\pm 4$ & $200\pm 10$ \\[0.5ex] \hline
\end{tabular}%
\end{table}

It is significant that the phenomenological approach with Eq.~(\ref{eq-delta}) can be applied only down to 70 K, where the cross-over in susceptibility has been observed and where the tetramer hamiltonian, Eq.~(\ref{eq-Hamil}), becomes insufficient. In the case of CuGeO$_3$ the increase of the line-widths close to the spin-Peierls transition (1 - 2 K) was attributed to the freezing of triplets, weakening the exchange narrowing~\cite{Eremina2003}. SeCuO$_3$, on the other hand, shows a line-width minimum around 50 K, much above $T_N$. Also, at the same temperature the crossing of two line-width curves occurs. Similarly broad minimum and subsequent increase of $\Delta H$ at lower temperatures has been observed in BaCu$_2$Ge$_2$O$_7$, another $S = 1/2$ AFM Heisenberg chain~\cite{Bertaina2004} which does not enter the spin-Peierls phase but orders at $\approx $ 8 K.

We also mention the contribution from the anisotropic Zeeman interaction to the ESR line-width, due to the presence of two inequivalent $g$-tensors. It has been shown~\cite{Pilawa1997,Heinrich2003} that in the high temperature limit that contribution is proportional to $(\Delta g)^2$. However, in $\sco$ the weakest coupling within the tetramer is $J_{12} \sim 160$ K and the high temperature limit refers to values much above our experimental window. To estimate the contribution from the anisotropic Zeeman interaction to $\Delta H$ one needs to know all the components of $\tilde{g}_1$ and $\tilde{g}_2$ so further investigation is desired.

%
%
%
%

\section{Discussion}
\label{Discussion}

Let us first discuss the possible origins of the 70 K cross-over. As mentioned in Sec.~\ref{sec-magnetization}, no combination of parameters in Eq.~(\ref{eq-chi}) can reproduce the susceptibility in the range 10 K $ < T < $ 70 K. Considering the structural arguments presented in Sec.~\ref{sec-structure}, it is natural to assume that the cross-over is a consequence of the inter-tetramer interaction along the $a$-axis, $J_{22}$. It is expected that this interaction is rather weak compared to $J_{11}$ and $J_{12}$ due to the large elongation of the Cu2 octahedron ($d$(Cu2 - O4) = 2.71 \AA). However, since the octahedron is not ideal, there maybe some mixing of $d_{x^2-y^2}$ and $d_{3z^2-r^2}$ orbitals in the Cu2 -- O4 coupling, allowing for a finite $J_{22}$. The value of the angle $\measuredangle $(Cu2 - O4 - Cu2) = 95$^0$ also suggests very weak $J_{22}$ and even allows for both FM and AFM couplings to be realized. Such a scenario would mean that below 70 K SeCuO$_3$ should be regarded as a system of magnetic chains composed of three different couplings with very disproportionate values.

Since the strength of $J_{22}$ sensitively depends on the exact values of $d$(Cu2 -- O4) and $\measuredangle $(Cu2 -- O4 -- Cu2), one cannot disregard the possibility that a small structural change, particularly a shift of the ligand position(s), may occur around 70 K and enhance $J_{22}$. One possibility is that $J_{22}$ becomes (more) ferromagnetic, which can come from the decrease of the $\measuredangle $(Cu2 -- O4 -- Cu2) angle towards 90$^0$, giving rise to an increase of the susceptibility relative to the bare tetramer picture. Such a change can come from a lateral shift of the O4 apical ion relative to the $d_{x^2-y^2}$ orbital, thus pulling the neighborig tetramers closer along the $a$-axis. However, this also reduces the distortion of the Cu2 octahedron which implies a smaller contribution from active $d_{x^2-y^2}$ orbitals and consequently a smaller $J_{22}$. In order to clarify this issue, further structural and theoretical studies are needed.

In addition to isotropic exchange interactions one needs also to consider the anisotropic ones, especially the Dzyaloshinskii-Moriya interaction (DMI). DMI mixes the singlet and the triplet states of localized spin clusters and is known to significantly influence the behavior of low dimensional magnetic systems~\cite{Choukroun2001,Herak2011}. Taking into account the symmetry considerations, DMI is allowed only for the Cu1 -- Cu2 pair. The main influence of DMI is to cant the neighboring moments which produces the ferromagnetic component. However, due to the inversion symmetry in the center of the tetramer, it is expected that the ferromagnetic contributions from each end cancel each other.

It is important to emphasize that the temperature dependence of $g$-factors and line-widths is not influenced by the feature around 70 K, indicating that the microscopic behavior is not greatly affected. 

Secondly, we discuss the observed rotation of magnetic axes, presented in Fig.~\ref{fig-torque}c. The observed phase shift is a result of the rotation of local magnetic axes, i.e. the change of the total $g$-tensor with temperature. Several origins can be considered. One could be the change of the ligand structure around the magnetic ion caused by a structural change. This type of change of the $g$-tensor was observed in CuSb$_2$O$_6$ where the temperature dependence of the $g$-tensor was triggered by structural changes in the wide temperature range \cite{Heinrich2003}. However, our structural analysis down to 80 K does not indicate any significant structural alterations.

Another explanation could be that the $g$-shift occurs due to present anisotropies induced by the increase of short range correlations at temperatures $T<J$ in low dimensional systems. Such $g$-shifts have been observed in the one-dimensional system Cu-benzoate where the temperature dependence and a rotation of the $g$-tensor axis has been observed \cite{Pilawa2001}. This was later explained as a consequence of the Dzyaloshinsky-Moriya interaction (DMI) within the Oshikawa-Affleck (OA) theory of ESR in spin $S=1/2$ Heisenberg chains \cite{Oshikawa2002,Oshikawa2007}. Similar $g$-shift was documented and explained within the OA theory in 1D Cu pyrimidine dinitrate \cite{Zvyagin2005}. However, in $\sco$ the $g$-shift saturates below 50~K only to increase again in the vicinity of the phase transition to long range order. This saturation is not expected if the $g$-shift is due to strengthening of anisotropic correlations.

Finally, a third, new scenario, which we argue is realized in $\sco$, can emerge in systems where there are at least two inequivalent sites contributing to the macroscopic anisotropy with different local magnetic axes and different temperature dependencies of magnetic susceptibility. In $\sco$ there are two copper sites, Cu1 and Cu2, with different values and orientations of their $g$-tensors. Since they are exchange-coupled, the ESR experiment observes only one line with a total $g$-tensor which non-trivially depends on $g_1$, $g_2$, $J_{11}$ and $J_{12}$ (to a lesser extent even on $J_{22}$). From the present set of data we are not able to fully model the system. In what follows we present a simple approach which depicts the essential nature of the physical processes.

Let us assume that individual tensors $\tilde{g}_1$ and $\tilde{g}_2$ are diagonal in a coordinate system defined by $ \{[10\bar{1}],b,\tau \} $. Within our tetramer model we set $J_{12} = J_{22} = J' = 0$, $J_{11} > 0$ and we observe how the total $g$-tensor depends on the temperature as the central Cu1 -- Cu1 pair experiences the singlet-triplet transformation. We use a simple averaging to get the total $g$-tensor
\begin{equation}
\tilde{g}_{tot} = \frac{\alpha \tilde{g}_1 + \beta \tilde{g}_2}{\alpha + \beta}
\label{eq-average}
\end{equation}
where $\alpha = \alpha_{T}/\alpha_{max}$, $\alpha_{max} = 3/4$, $\beta = 1$ and $\alpha_{T}$ is the probability of finding the Cu1 -- Cu1 dimer in a triplet state
\begin{equation}
\alpha_{T} = \frac{3 e^{-J/T}}{1 + 3 e^{-J/T}} \text{.}
\label{eq-triplet}
\end{equation}

In Fig.~\ref{fig-gfactor} we present the results of the above procedure for the $[10\bar{1}]$- and $b$-directions using the values $\tilde{g}_1 = (2.335,2.14,2.07)$, $\tilde{g}_2 = (2.153,2.33,2.07)$ and $J_{11}$ = 290 K. The agreement with the experimentally obtained values is surprisingly good, given the simplicity of the model, and the obtained value of $J_{11}$ is very close to the one extracted from Eq.~(\ref{eq-chi}).

\begin{figure}
\includegraphics[width=0.45\textwidth]{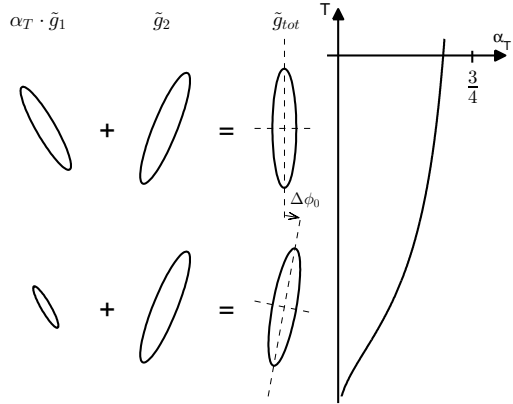}
\caption{Averaging the $g$-tensors with varying contribution from $\tilde{g}_1$ which is proportional to the triplet probability (the graph on the right-hand side). As the temperature is lowered, the total $g$-tensor effectively rotates by $\Delta \phi_0$. The semi-axis values are not scaled with the values from the text.}
\label{fig-ellipse}
\end{figure}

Within the choices of model parameters, it is not possible to describe the torque: no rotation of magnetic axes occurs since both g-tensors are diagonal in the same coordinate system. A more realistic approach would take into account two $g$-tensors for each copper site whose orientation is set by the local ligand environment. Since in that case the principal axes for $\tilde{g}_1$ and $\tilde{g}_2$ are pointing in different directions, the reduction of the Cu1 moment will also influence the direction of principal axes of the total $g$-tensor. In Fig.~\ref{fig-ellipse} we demonstrate such a scenario for two temperatures with the emphasis on the rotation of the magnetic axes of the total $g$-tensor as observed with torque measurements (see Fig.~\ref{fig-torque}c).

In $\sco$ this effect is not averaged out by the symmetry related sites. If the g-tensor is diagonal in the experimental coordinates, the diagonal components remain dominant also in the crystal coordinate system. By using the symmetry operators of the space group of the crystal, it is straightforward to show that the g-tensors at the symmetry-related sites within the tetramer do not cancel out: actually the dominant diagonal values add up. Also, the screw transformation maps tetramers from one chain to the other. Since $g$-tensors are invariant under the 180$^0$ rotation, the contributions from 4 different Cu1 and Cu2 sites in the unit cell are adding up.

This simplified model demonstrates how effective magnetic axes can rotate when inequivalent sites experience different quantum correlations. However, to quantitatively fit our data, one would need the complete Hamiltonian, a theoretical method to treat it and a theory relating the calculated site dependent susceptibilities to the observed ESR signal.

%
%
%
%
%

\section{Conclusions}
\label{Conclusion}

We have demonstrated that the tetramer model of $\sco$ can provide a good explanation for the observed high temperature behavior of the susceptibility and especially of the unusual temperature dependence of the $g$-factor. The kink in the susceptibility around 70 K indicates that the tetramer hamiltonian does not provide a complete set of interactions in the system. An additional coupling, most probably $J_{22}$ which couples tetramers along the $a$-axis, also influences the line-width of the ESR spectra. Below 70 K the line-widths develop a minimum and the overall behavior is similar to the case of a 1D $S=1/2$ AFM Heisenberg chain BaCu$_2$Ge$_2$O$_7$, although $\sco$ shows a large difference between various couplings ($J_{11} > 70$ K).

The inter-tetramer coupling, $J'$, which is mediated via Cu -- O -- Se -- O -- Cu paths, drives the system into a long-range ordered state at $T_N = 8$ K. This provides a wide temperature window where the effects of inter-tetramer quantum fluctuations can be investigated. Compared to the incommensurate compounds Cu$_2$Te$_2$O$_5X_2$, $\sco$ shows a simpler magnetic structure with a relatively small spin-flop field $\sim 1.7$ T. This can enable an easier theoretical modelling of the magnetic structure and magnetic excitation with and without the magnetic field.

Finally, we have argued that the observed temperature dependence of magnetic axes is due to the inequivalent copper sites experiencing different type of quantum correlations. Assuming temperature independent $g$-tensors, which is valid in most compounds, this effect can be inverted to a new experimental method of obtaining site selective susceptibilities from bulk measurements.

%
%
%
%
%

\section{Acknowledgement}
\label{Acknowledgement}

We acknowledge the financial support from Projects No.~035-0352843-2845, No.~035-0352843-2846, No.~119-1191458-1017 and No.~035-0352827-2842 of the Croatian Ministry of Science, Education and Sport, 02.05/33 of the Croatian Science Foundation and the Swiss National Science Foundation and its NCCR MaNEP.

\end{document}